# 10 YEARS OF WIRE EXCITATION EXPERIMENTS IN THE CERN SPS

F. Zimmermann, CERN, Geneva, Switzerland

*Abstract*

This paper reviews the set-up, experimental studies, and beam observations with one or two prototype long-range beam-beam 'wire' compensators in the CERN SPS from 2002 to 2012.[*]

## MOTIVATION

Following earlier studies investigating the effect of long-range collisions for the SSC [1] and LHC [2, 3], in 1999 weak-strong beam-beam simulations for the LHC – using the modelling recipe of Ref. [4] – revealed the existence of a diffusive aperture at transverse amplitude of 6-7$\sigma$, which is induced by the nominal long-range beam-beam encounters [5]. An example simulation result is illustrated in Fig. 1.

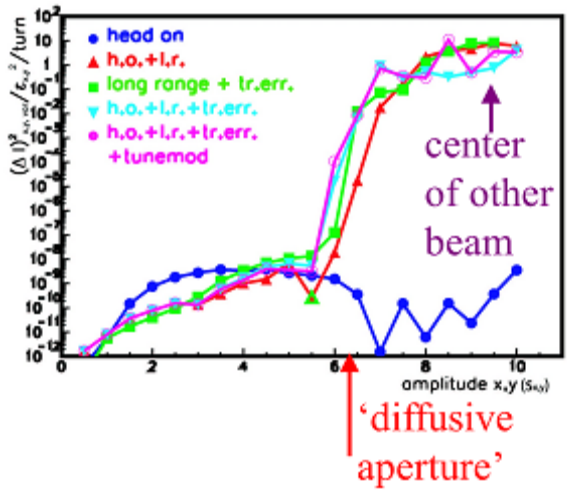

Figure 1: Transverse action diffusion rate $\Delta I^2_{rms}/\varepsilon^2_{x,y}$/turn as a function of transverse amplitude in units of $\sigma$ under various conditions obtained from a weak-strong beam-beam simulation [5].

## COMPENSATION SCHEME

The simulated strong effect of the LHC long-range collisions inspired the search for mitigation, and in 2000 J.-P. Koutchouk proposed a long-range beam-beam compensation for the LHC based on current-carrying wires [6]. At a transverse distance, the wires generate the same transverse force of shape $1/r$, as the field of the opposing beam at the parasitic long-range encounters [6]. In order to correct all non-linear effects the correction must be local. For this reason, there needs to be at least one wire compensator, in the CERN internal naming convention called 'BBLR', on one side of each primary interaction point (IP) for either beam, in a region where the two beams are already physically separated, but otherwise as close as possible to the common region where the long-range encounters occur. The proposed layout features the compensator 41 m upstream of the separation dipole D1, on both sides of IP1 and IP5, where the horizontal and vertical function are equal, as is shown in Fig. 2. Figure 3 illustrates how one wire cancels the effect of all 16 long-range encounters occurring on one side of the IP. The betatron phase difference between the BBLR and the average LR collision is 2.6° (ideally it should be zero).

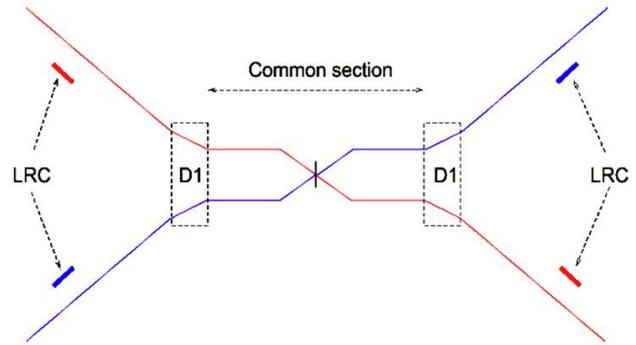

Figure 2: Schematic location of proposed LHC wire compensators [6, 7].

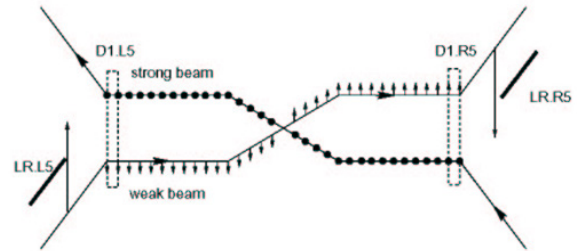

Figure 3: Illustration of the compensation principle [6, 7].

In simulations the wire compensator was shown to effectively shrink the tune spread caused by the long-range collisions to essentially zero [6] (Fig. 4) and to gain about 1.5$\sigma$ in diffusive aperture [8] (Fig. 5).

Strong-strong beam-beam simulations including wire compensators were reported in Ref. [9], and further analytical studies of the onset of chaos due to the long-range collisions in Ref. [10].

## SPS WIRE COMPENSATORS

In order to explore the 'simulated' effect of long-range encounters and to benchmark the simulations with the SPS beam, in 2002 a first prototype compensator was fabricated and installed. This BBLR consisted of two 80-cm long units (each with a wire length equal to 60 cm), installed one behind the other, and each containing a single water-cooled wire, vertically displaced from the beam centre. Two years later, a second BBLR was constructed, equipped with three wires of different transverse orientation. The second BBLR also consisted of two units of the same length, like the first one, but mounted on a movable support so that their vertical position could be varied over a range of 5 mm through remote control. A primary purpose of this second BBLR, installed at a betatron phase advance of about 3° from the first one (hence similar to the phase advance between the proposed location of the LHC wire compensator and the centre of the long-range collisions), was to demonstrate the effectiveness of a realistic compensation scheme, which could be simulated by powering the vertical wire(s) of the two BBLRs with opposite polarity. In addition to the vertical wire, a horizontal wire and a wire at 45° were added to allow for experimental studies and comparisons of various crossing schemes (horizontal-vertical, vertical-vertical, and 45°).

---

[*] This work was supported, in parts, by the European Commission under the FP7 Research Infrastructures project EuCARD, grant agreement no. 227579.

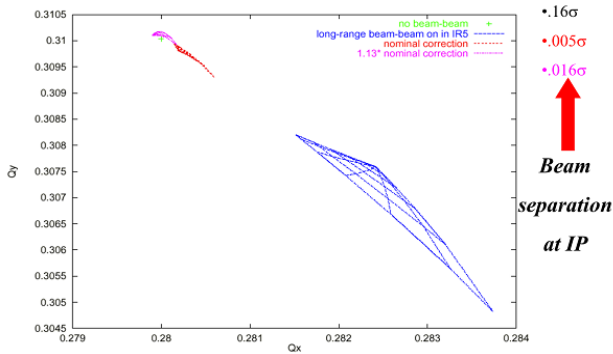

Figure 4: Simulated LHC tune footprint due to long-range collisions with and without wire compensator [6].

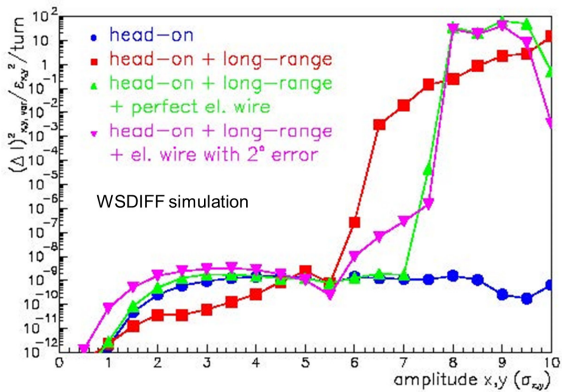

Figure 5: Simulated LHC diffusive aperture with ideal (green) and realistic wire compensator (pink) compared with the case of no compensation (red) and head-on collisions only (blue) [8].

Photographs of both devices are shown in Fig. 6, as well as technical drawings in Figs. 7 and 8. The wire of the first BBLR is mounted at a fixed nominal vertical distance of 19 mm from the centre of the chamber (so that it is in the shadow of the SPS arc aperture). More details and documentation on the SPS wire compensator prototypes (and the experiments conducted in the SPS using these devices) can be found on a dedicated web site [11].

The needed wire current $I_w$ is related to the number of long range collisions to be compensated, the length of the $l_w$ and the bunch population $N_b$, as $I_w = N_b ec \#LR / l_w$, where $e$ denotes the elementary charge and $c$ the speed of light. The two 60-cm long wires of one unit can be excited with up to 267 A of current, which, according to the above equation, produces an effect equivalent to 60 LHC LR collisions (e.g., roughly the combined effect of all nominal long-range encounters around IPs 1 and 5).

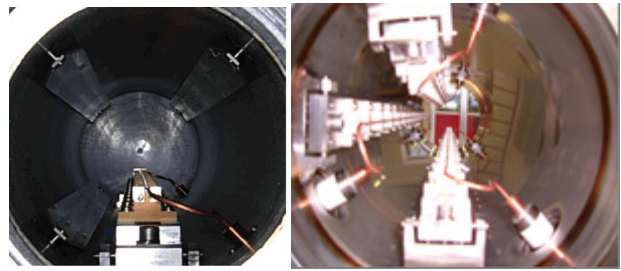

Figure 6: The first (left) and the second prototype wire compensator (right) installed in the CERN SPS in 2002 and 2004, respectively.

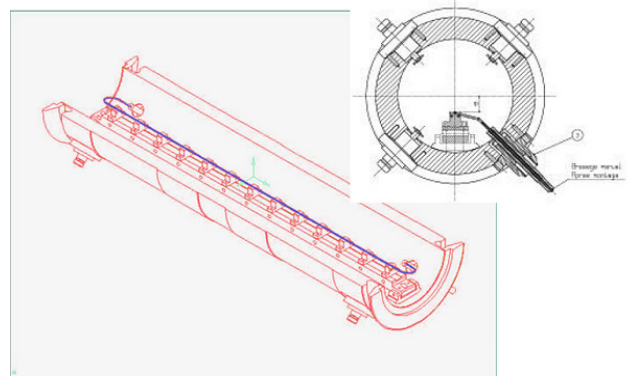

Figure 7: Technical drawings of the first SPS wire compensator (2002).

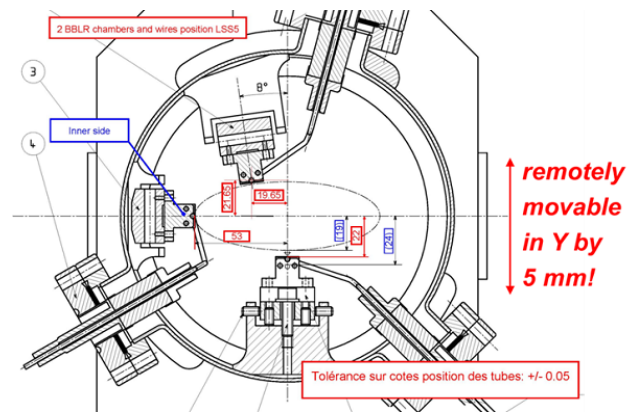

Figure 8: Technical drawing of the second SPS wire compensator (2004).

Figure 9 presents a side view of the first BBLR device. Each BBLR, consisting of two units, has a total length of (2 × 0.8 + 0.25) m = 1.85 m. A photograph shows BBLRs 1 and 2 installed in the SPS tunnel (Fig. 10). Figure 11 illustrates the horizontal and vertical beta functions along the two × two BBLR units. The average value of the beta functions is about 50 m.

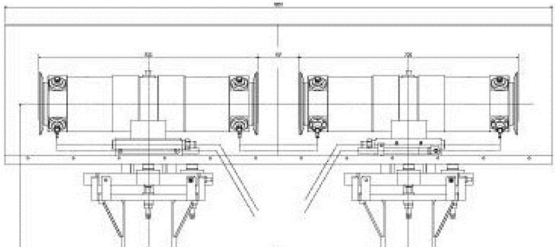

Figure 9: Side view of SPS BBLR #1.

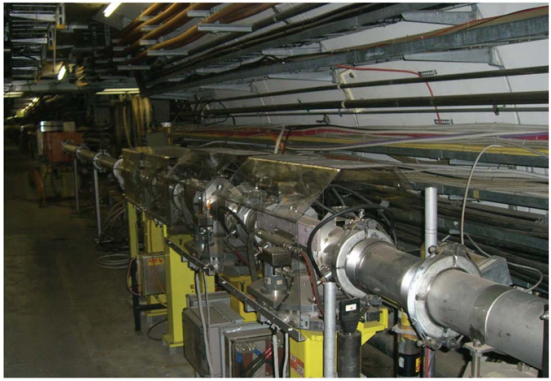

Figure 10: SPS BBLRs no. 1 and 2 (4 boxes) installed in SPS Straight Section 5.

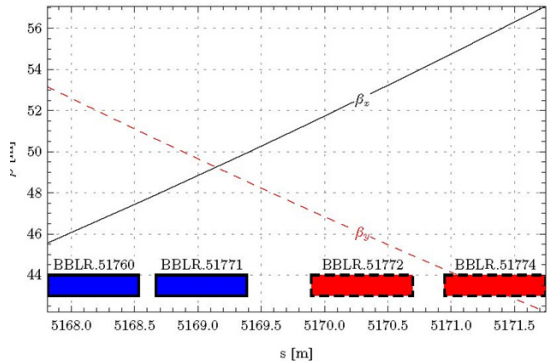

Figure 11: Horizontal and vertical beta functions across the two SPS BBLRs (each consisting of two units).

Additional compensator wire units are available at CERN. A complete BBLR consisting of two units with water-cooling, similar to BBLR no. 2, is ready (repaired after an earlier leak). Two air-cooled BBLRs from the Relativistic Heavy Ion Collider (RHIC) have been shipped from Brookhaven National Laboratory and are in store at CERN [12]. Thus, including the two BBLRs presently installed in the SPS, a total of five sets are (or have been) available.

## SCALING LAWS

The perturbation by the wire compensator at distance $d$ from the beam centre is

$$\Delta y' = \frac{2 r_p l_w I_w}{\gamma e c (y-d)} ,$$

from which the relative perturbation at the dynamic aperture becomes

$$\frac{\Delta y'}{\sigma_{y'}} = \left(\frac{2 r_p l_w}{ec}\right)\left(\frac{I_w}{(\gamma \varepsilon)\tilde{n}_{da}}\right) ,$$

where $\tilde{n}_{da}$ denotes the dynamic aperture in units of the rms beam size, $l_w$ the wire length and $I_w$ the wire current. This equation shows that, for constant normalized emittance, the effect in units of sigma is independent of energy and beta function. In *scaled experiments* the wire current is varied in direct proportion to the factor by which the emittance differs from the desired emittance.

## HISTORY OF SPS BBLR STUDIES

The SPS BBLRs were used to perform the following beam studies:

- perturbation by single wire as LHC LR simulator (2002 to 2003) [13,14];
- two wire compensation, scaled experiments, distance scan (2004) [15,16];
- tests of crossing schemes (2004) [15,16,17];
- one and two wires at different energies: 26, 37, and 55 GeV/$c$; scans of $Q'$, distance, current (2007) [18,19,20];
- two-wire compensation with varying $Q$, $I_w$, $Q'$ scans at 55 GeV/$c$ (2008) [21,22];
- two-wire compensation and excitation in coasts at 120 GeV/$c$ (2009) [22]; and
- two-wire compensation and excitation in coasts at 55 GeV/$c$ (2010) [23].

Figure 12 illustrates typical SPS cycles used towards the end of the last decade for BBLR studies at three different beam energies. During dedicated machine studies, at the target energy the SPS cycle could be stopped and the beam be made to 'coast' for e.g. ten minutes for measurements of the beam lifetime in steady-state conditions and parameter scans.

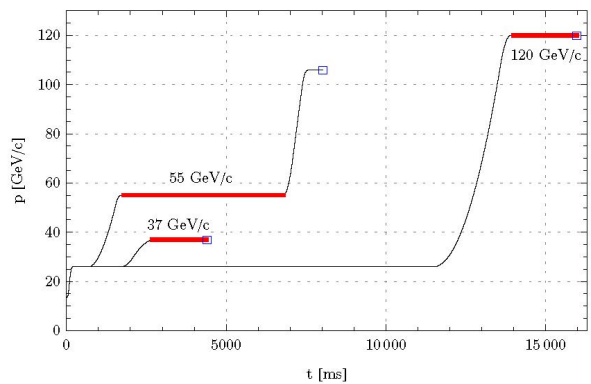

Figure 12: SPS cycles during experiments in 2008 and 2009 [G. Sterbini].

## TECHNICAL ISSUESS

A number of technical issues had to be addressed, especially in the early days of the SPS BBLR studies. These included:

- installation of dedicated ion chambers and photomultiplier tubes (PMTs) near the BBLR;
- the addition of an inductive coil to suppress wire-current ripple;
- computation and experimental verification of wire heating;
- emittance blow-up by means of the transverse damper or by injection mismatch together with resonance crossing (equalizing the vertical and horizontal emittances) so as to achieve the nominal LHC parameters or to increase sensitivity;
- use of fast wire scanners and scrapers;
- installation of a dedicated dipole near the BBLR to correct the induced orbit change locally;
- continuous tune corrections;
- preparation and use of multiple superimposed orbit bumps to vary the beam-wire distance;
- (later) choice of higher beam energy: 37, 55 or 120 GeV/$c$ (for good lifetime without wire excitation); and
- (later) experiments in coast (to avoid transient data).

Figure 13 illustrates the combination of orbit-corrector bumps used to vary the beam-wire distance at higher beam energy. The resulting minimum normalized distance in units of rms beam size depends on the beam energy and on the normalized emittance as shown in Fig. 14. In case the emittance was too small the beam could be blown-up with transverse feedback and resonance crossing.

The natural SPS beam lifetime was about 30 h at 55 GeV/$c$, but only 5-10 min at 26 GeV/$c$ (where the physical aperture was only about 4$\sigma$).

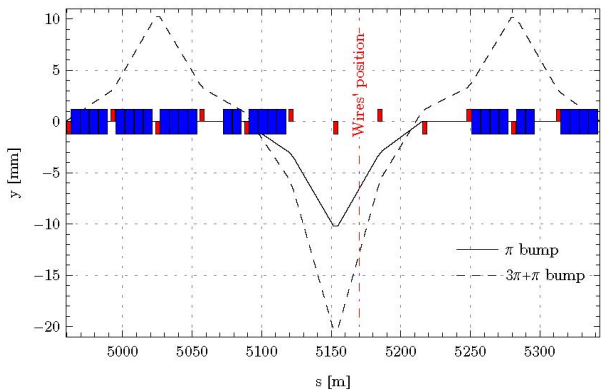

Figure 13: Superimposed 3+5 corrector bumps at the SPS wire compensator [G. Sterbini].

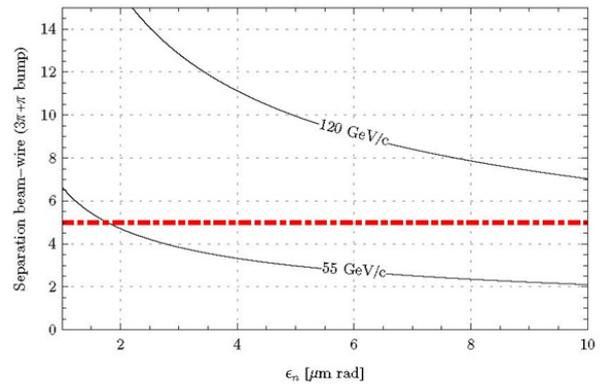

Figure 14: Minimum normalized distance in units of rms beam size as a function of normalized emittance for two beam energies [G. Sterbini].

## SINGLE BBLR 'EXCITATION' STUDIES

Changes in orbit and tunes allow for a precise determination of the beam-wire distance. Example data from 2002 are shown in Figs. 15-17.

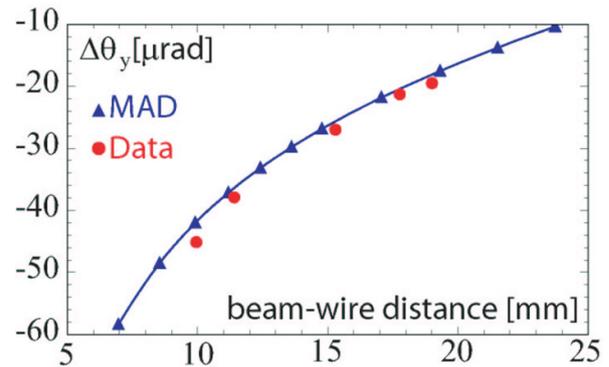

Figure 15: Deflection angle at the wire compensator as a function of beam-wire distance, comparing data and measurements [14] [J. Wenninger].

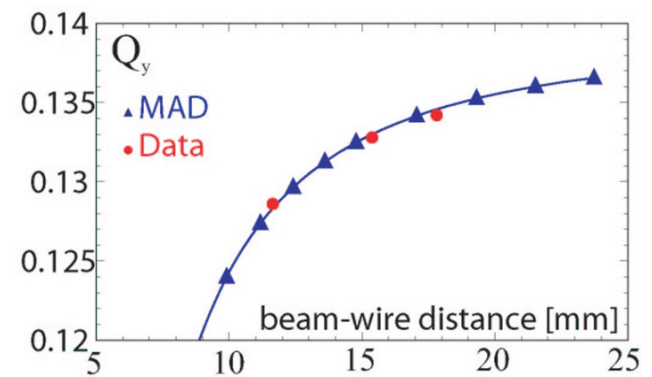

Figure 16: Vertical tune change as a function of beam-wire distance, comparing data and measurements [14] [J. Wenninger].

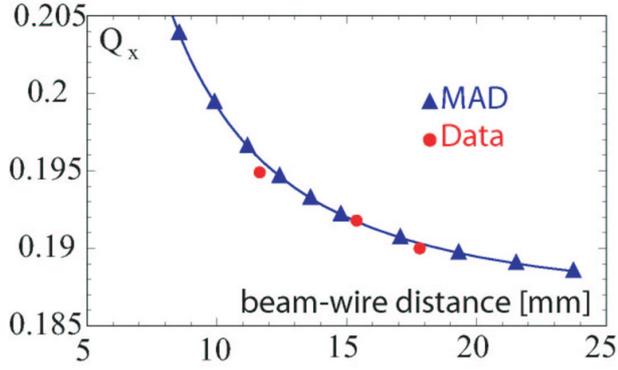

Figure 17: Horizontal tune change as a function of beam-wire distance, comparing data and measurements [14] [J. Wenninger].

The change in the beam orbit at the compensator follows from the self-consistent equation

$$\Delta d = \frac{\beta_y I_{wire} l_{wire} r_p}{\gamma ec(d+\Delta d)\tan(\pi(Q_y+\Delta Q_y))},$$

while the tune changes are given by

$$\Delta Q_{x,y} = \mp \frac{r_p \beta_{x,y} I_{wire} l_{wire}}{2\pi\gamma ec} \frac{1}{(d+\Delta d)^2}.$$

In most of the later studies only the tune change was monitored.

The effect of the compensator on the nonlinear optics has also been studied, by acquiring turn-by-turn beam-position monitor (BPM) data after kicking the beam. The nonlinearity of the wire field led to a reduced decoherence time, to a tune shift with amplitude, and (additional) spectral resonance lines. The measured tune shift was consistent with the theoretical predictions

$$\Delta Q_x \approx \frac{3}{4} \frac{I_{wire} l_{wire} r_p}{\gamma ec} \frac{\beta_x}{d^4} \hat{y}^2$$

and

$$\Delta Q_y \approx -\frac{3}{8} \frac{I_{wire} l_{wire} r_p}{\gamma ec} \frac{\beta_x}{d^4} \hat{y}^2$$

The resonance lines introduced by the BBLR are illustrated in Fig. 18.

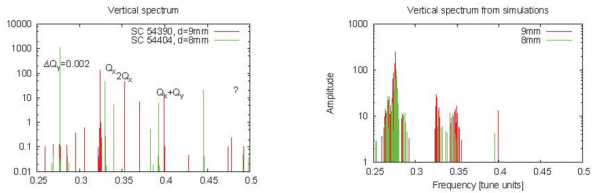

Figure 18: Resonance spectra with wire excitation: experimental data with 240 A wire current at 9 (red) and 8 mm (green) beam-wire distance (left) and the corresponding simulation data (right) [18] [U. Dorda].

A strong effect of chromaticity was noticed when the compensator was excited. Figure 19 shows the beam intensity evolution during $Q_{x,y}'$ scans at 37 GeV/$c$.

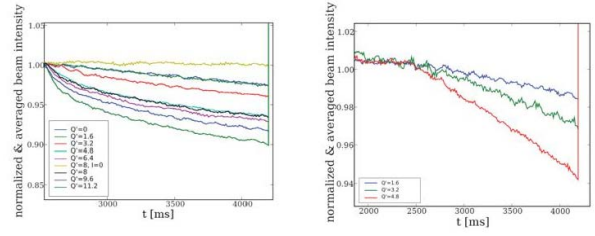

Figure 19: Beam intensity as a function of time for various values of the horizontal ($Q_x'$, left) or vertical chromaticity ($Q_y'$, right) [20]. The wire excitation was 180 A-m, the beam momentum 37 GeV/$c$ and the normalized beam-wire separation about 6.5σ (9 mm).

Figure 20 compares the measured (left) and simulated beam loss (right) for two different values of the vertical chromaticity as a function of the integrated wire strength.

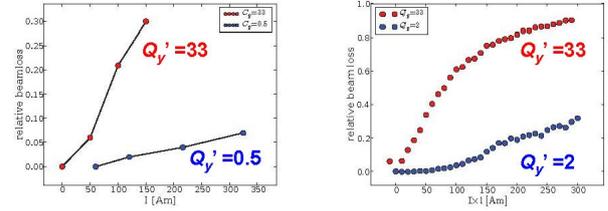

Figure 20: Relative beam loss for two different values of the vertical chromaticity as a function of wire excitation in units of A-m, comparing experimental data (left) and simulations (right) [20] [U. Dorda]. The beam-wire separation was ~6.6 σ.

Various attempts were made to directly measure the 'diffusive' or dynamic aperture. To this end, three types of signals were used: (1) lifetime and background, (2) beam profiles and final emittance, and (3) local diffusion rate inferred by scraper-retraction experiments. Figures 21 and 22 shows some example measurements of lifetime and background at 55 GeV/$c$. A drop in the lifetime and increased losses are observed for separations less than 9σ; at 7-8σ separation the lifetime decreases to 1-5 h. These results indicated that the LHC nominal separation of 9.5σ (for the encounters between the IP and the first quadrupole Q1) is well chosen, but 'close to the edge'.

Beam profiles before and after wire excitation, measured with an SPS 'wire scanner' (fully unrelated to the wire compensator), reveal that the particles at large transverse amplitude are lost due to the wire excitation; see Fig. 23. These measurements confirmed that the wire compensator or the equivalent set of long-range encounters, acts as a highly effective scraper.

This type of measurement allows for an estimate of the diffusive/dynamic aperture. Specifically, an Abel transformation of the wire-scan data of the form [23, 24]

$$\rho(A) = -2A \int_A^R d\eta \frac{g'(\eta)}{\sqrt{\eta^2 - A^2}}$$

can be used to compute the change in the (normalized) amplitude distribution due to the wire excitation. For the data of Fig. 23 the results are presented in Fig. 24, indicating that in this particular example (with intentionally small separation) the dynamic aperture is at about 1 σ).

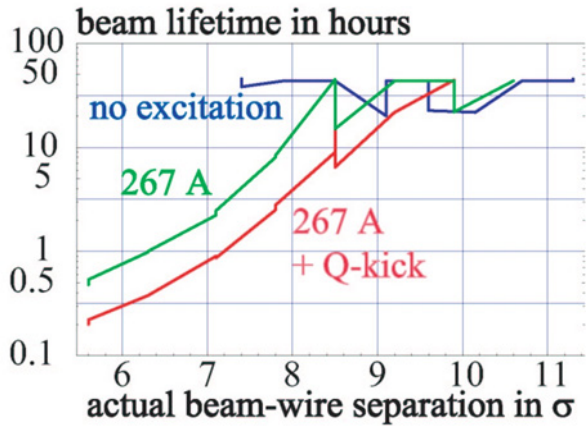

Figure 21: Lifetime as a function of the wire-beam separation in units of rms beam size with (green) and without (blue) wire excitation at 267 A, which corresponds to the nominal total number of LHC long-range encounters at IPs1 and 5. The red data were also taken with the wire excited, while in addition firing the (weak) tune kicker to add a further perturbation.

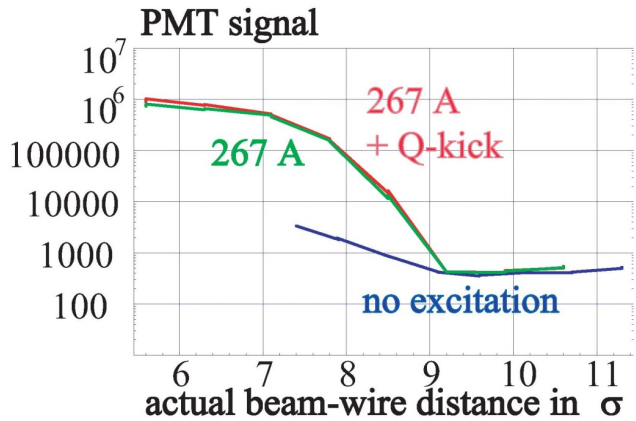

Figure 22: Local relative beam loss rate measured by a photomultiplier as a function of the wire-beam separation in units of rms beam size with (green) and without (blue) a wire excitation of 267 A. As in Fig. 21 for the red data set the (weak) tune kicker was repeatedly fired while the wire was excited.

Figure 25 displays the final emittance inferred from the beam profiles as a function of beam-wire distance without wire excitation and for wire currents of 67 A and 267 A (the latter corresponding to 60 LHC long-range encounters). The reduction of the final emittance without wire excitation at smaller distances is due to mechanical scraping of the beam by the edge of the wire.

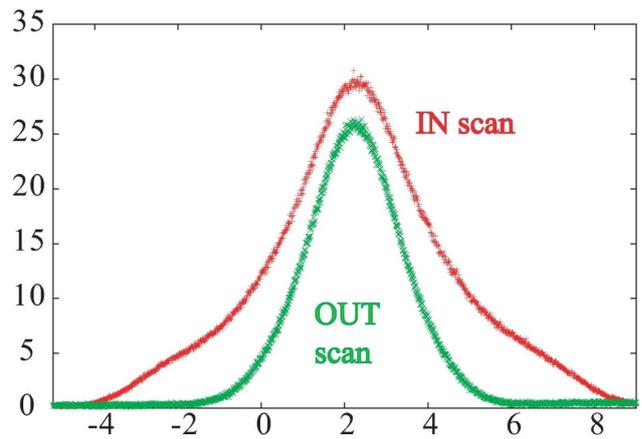

Figure 23: Beam profile before and after wire excitation measured at 26 GeV/$c$. The inferred initial and final emittances were 3.40 μm and 1.15 μm, respectively.

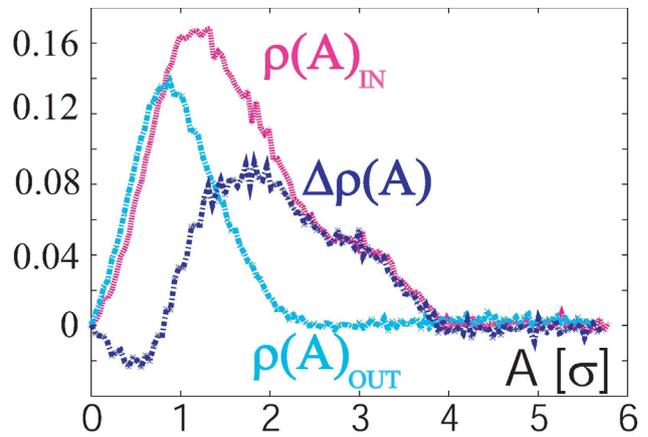

Figure 24: Abel transformation of the beam-profile data from Fig. 23, revealing the change in the (normalized) amplitude distribution.

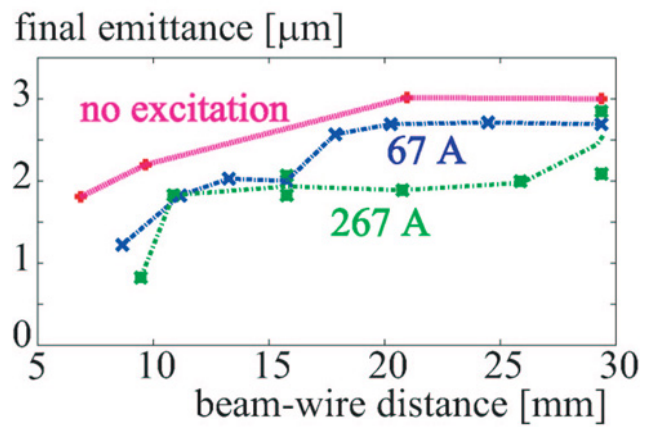

Figure 25: Final emittance without (red) and with wire excitation (blue 67 A, green 267 A) at a beam momentum of 26 GeV/$c$ as a function of beam-compensator distance.

With the Abel-transformation technique it was not always possible to obtain a clean result for the diffusive aperture. Therefore, a different technique was also employed to infer the variation of the diffusive/dynamic aperture. Namely, without wire excitation, a known aperture restriction was introduced using a dedicated mechanical 'scraper,' and wire scans were then executed to determine the 'final emittance' corresponding to a given known aperture determined by the scraper position. This calibration measurement is presented in Fig. 26 – the curve of measured final emittance as a function of scraper position allows estimation of the effective aperture due to the wire excitation from the associated 'final emittance' value.

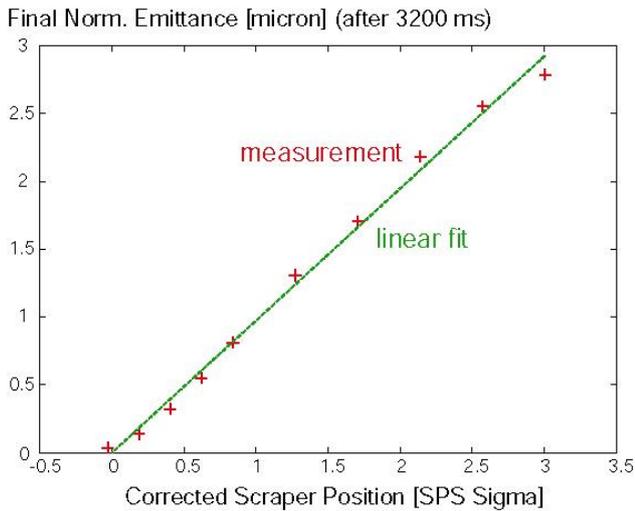

Figure 26: Calibration of the final emittance values by a mechanical scraper.

Following this plan and using the calibration curve of Fig. 26, measurement results for different wire currents were converted into normalized diffusive apertures. The result, shown in Fig. 27, suggests a linear dependence of the dynamic aperture on the square root of the wire current, which is consistent with a scaling law first pointed out by Irwin [4]. In the figure, the measured dynamic aperture is smaller than the simulated diffusive aperture, especially at lower current, hinting at additional effects not included in the simulations or at a systematic error in the calibration method.

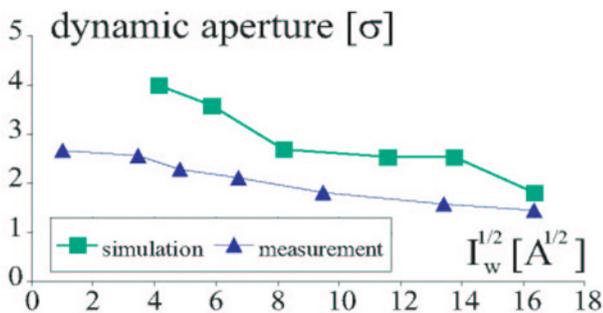

Figure 27: Effect of wire current on SPS dynamic aperture (26 GeV/$c$), inferred from final emittance and the calibration of Fig. 26.

Yet another approach to measuring the diffusive aperture is to directly detect the diffusion rates at various transverse amplitudes, by inserting a scraper to remove particles in a small area around the target amplitude article, then retracting this scraper by a small step, and observing how the loss signal reappears as particles diffuse outwards to the new position of the scraper. This type of measurement was previously used at HERA (and elsewhere) to determine the local diffusion coefficients [25]. Unfortunately, scraper retraction attempts for the SPS wire-compensator studies were not very successful. Figures 28 and 29 present example results. In Fig. 29 the scraper position is about $1\sigma$, and a temporary decrease in the loss rate can be noticed, which might be used to fit a diffusion constant. However, at larger amplitudes (which are of greater interest) the diffusion was much faster than the speed of the SPS scraper. Figure 29 also shows that the scraper moving to its target position, while the wire is not yet excited, already intercepts a significant halo, as evidenced by the elevated background rate prior to the wire excitation.

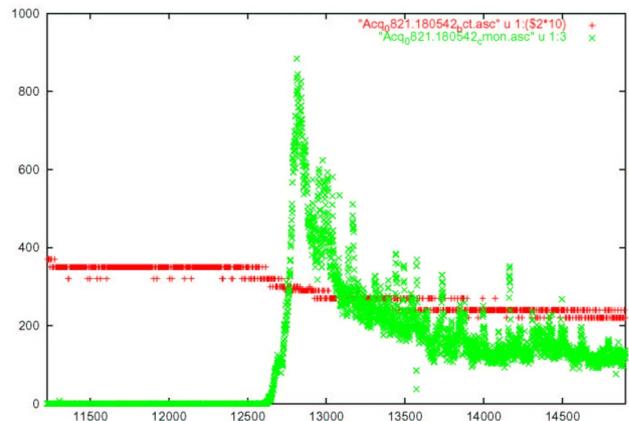

Figure 28: Beam current in units of $10^8$ protons (BCT, red) and local loss rate detected by photomultiplier (green) as a function of time during the cycle in units of ms. For these data the wire was excited (at 12725 ms) but no scraping was applied.

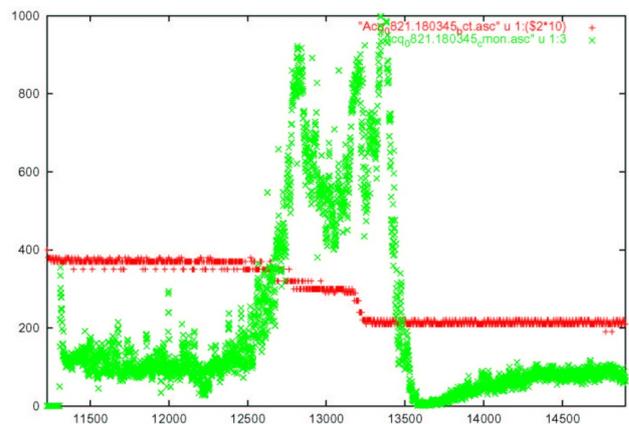

Figure 29: Beam current in units of $10^8$ protons (BCT, red) and local loss rate detected by photomultiplier (green) as a function of time during the cycle in units of ms. For these data the wire was excited at 12725 ms; later, at 13225 ms, the scraper was inserted and retracted.

One of the most interesting results from the SPS wire measurements is the measured dependence of the 'beam lifetime' $\tau_{beam}$, as inferred from the beam loss during a cycle at 26 GeV/$c$, on the beam-wire distance $d$ [15], illustrated in Fig. 30. The measured dependence extremely well follows a 5$^{th}$ order power law as seen from the fitting result embedded in the figure. It has been suggested [26] that a nearby low-order resonance of order $n$ should cause a dependence $\tau_{beam} \sim 1/d^{n+1}$ and that the power in the exponent should, therefore, depend on the betatron tunes. Indeed at the Tevatron (with an electron lens applied as 'wire') [27] and at RHIC [28], operating at other working points in the tune diagram, different power laws were observed (third power and linear dependence, respectively). Figure 31, presenting SPS data for three different sets of tunes, taken several years later at a higher energy, confirms that the losses due to the wire are strongly tune dependent.

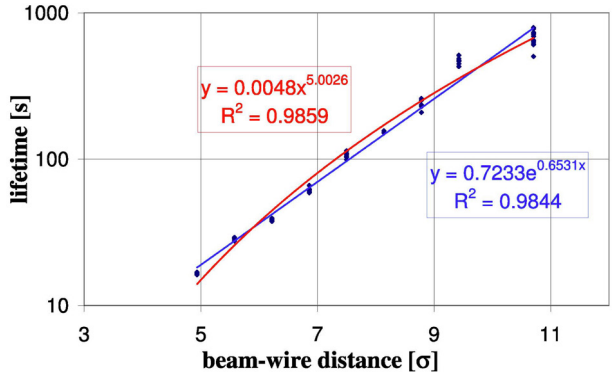

Figure 30: Beam lifetime as a function of beam-wire distance at 26 GeV/$c$, for betatron tunes of $Q_x = 0.321$ and $Q_y = 0.291$ [14].

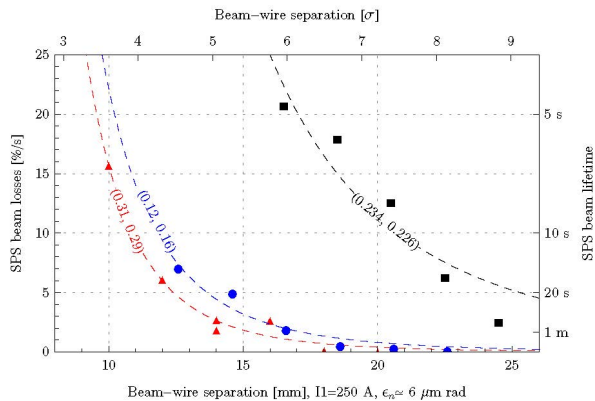

Figure 31: Beam losses as a function of beam-wire distance for three different pairs of tunes at 37 GeV/$c$ with 1.1 s cycle [21] [G. Sterbini]

Extrapolating this measurement to the nominal LHC beam-beam distance, $\sim 9.5\sigma$, predicts a 6 min lifetime. This result was one of the motivations for raising the SPS beam energy and for performing measurements with coasting (non-cycling machine) beams in later studies, where the beam lifetimes were significantly higher.

Figure 32 shows beam losses as a function of wire current $I_w$ for different normalized beam-wire separations (in units of σ) $d_n$. These later results were fitted as [21]:
$$\text{beam loss (\%)} = 0.07\, e^{-d_n} I_w^2 \,.$$

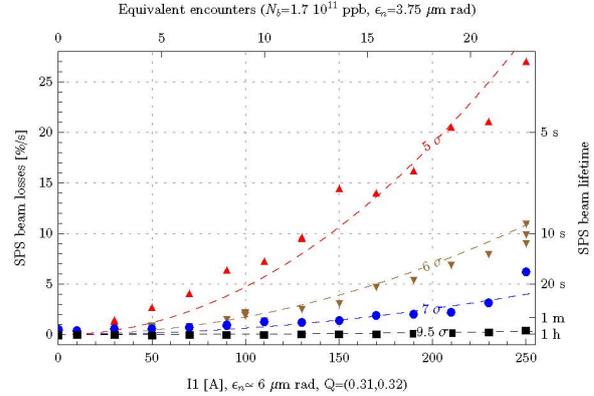

Figure 32: Beam losses as a function of wire current at 37 GeV/$c$ with a 1.1 s cycle, for betatron tunes of $Q_x = 0.31$ and $Q_y = 0.32$ (nominal values for LHC collisions) [21].

## COMPENSATION STUDIES WITH TWO BBLR WIRE COMPENSATORS

Experiments with two wire devices became possible after the installation of the second SPS wire compensator (Fig. 8) in 2004. The main focus of the two wire studies was the demonstration of compensating the effect of one wire by the second, about 3° apart in betatron phase advance from the first, and with a slightly different beta function ratio (Fig. 11), and a study of the associated tolerances, making use of the fact that the vertical position of the second wire can be controlled remotely over a 5 mm range. Such a study had been requested by the CERN LHC Technical Committee (LTC) in 2002. Other two-wire studies used the three independent wires of the second device (vertical, horizontal and 45° wires) to 'model' different crossing schemes at the two main interaction points of the LHC. Results from these latter studies are reported in the appendix.

Figure 33 shows a typical measurement result of the beam lifetime as a function of vertical tune on the SPS injection plateau, with the horizontal tune set to the nominal LHC collision value (0.31). The LHC vertical collision tune is 0.32, near the upper end of the scan range. Three cases are compared: no wire excitation, one wire excited, and both wires excited in compensating configuration. For this measurement the tune and orbit changes due to the wire compensation were corrected at each point. The data demonstrate that the lifetime reduction due to the first wire was recovered by the second wire over a large tune range, except at $Q_y < 0.285$ (close to the 7th-order resonance; see Fig. 34) or when approaching the third integer resonance (including at the nominal tune). Figure 35 shows a later result, taken at a higher beam energy (37 GeV/$c$), where the natural beam lifetime, with wires off, was much higher than at injection. Here the compensation worked well over an even larger range, but it still degraded close to the third integer resonance ($Q_y > 0.31$), and close to the 4th-order resonance ($Q_y < 0.27$).

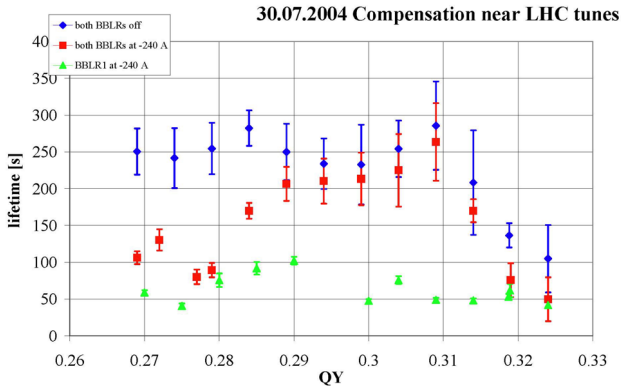

Figure 33: Beam lifetime as a function of vertical tune without wire (blue), with one wire excited at 240 A (green), and with both wires in compensating configuration (red) for a fixed horizontal tune of Qx = 0.31 [14]. The tune scan corresponds to the red line in Fig. 34. This measurement was performed in 2004 at a beam momentum of 26 GeV/c.

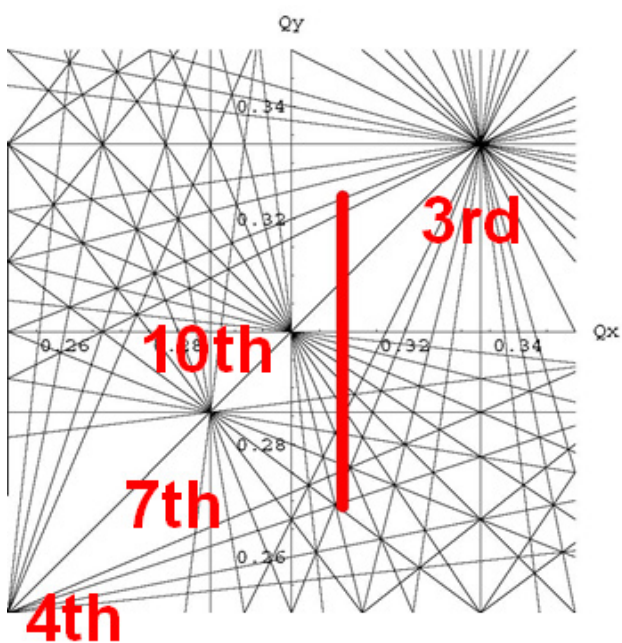

Figure 34: Tune diagram with some low-order resonance lines (incomplete), and the region of a typical vertical tune scan, as used, e.g. for the measurement of Fig. 33.

Figure 36 presents the results of a scan of the vertical position of the second wire compensator with respect to the (fixed) first wire, and the comparison with a simulation using the code BBSIM. The simulation predicts there to be no compensation beyond ~3 mm. The measurement revealed that the compensation was fully lost beyond ~2.5 mm from optimum (equivalent to ≤2σ).

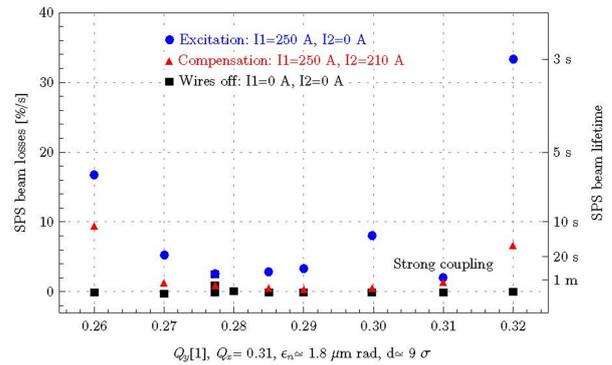

Figure 35: Beam losses (left) and beam lifetime (right) as a function of vertical tune without wire (black), with one wire excited at 250 A (blue), and with both wires in compensating configuration (red) for a fixed horizontal tune of $Q_x = 0.31$ [21]. The tune scan corresponds to the red line in Fig. 34. This measurement was performed in 2008 at a beam momentum of 37 GeV/c over 1.1 s.

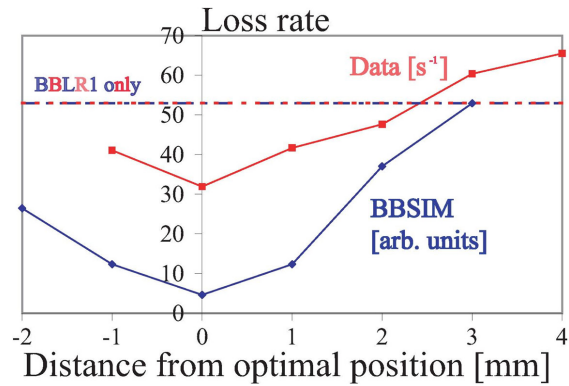

Figure 36: Beam loss rate as a function of vertical distance of second wire with respect to optimum location compared with BBSIM simulations [15].

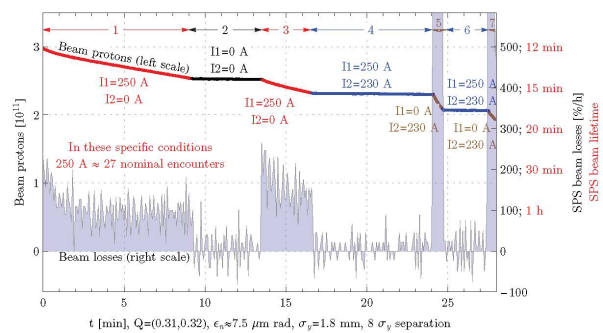

Figure 37: Beam intensity (left), beam loss and lifetime (right) as a function of time while one or two wire compensators at 8σ distance were either turned on (at 250 A and lifetime-optimized 230 A, respectively) or off, in coast at 120 GeV/c (2009) [21]. The normalized transverse emittance had intentionally been blown up to 7.5 μm (twice the nominal).

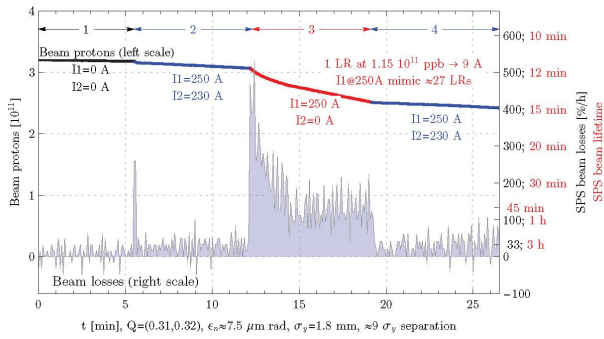

Figure 38: Beam intensity (left), beam loss and lifetime (right) as a function of time while one or two wire compensators at 8σ distance were turned on (at 250 A and lifetime-optimized 230 A, respectively) and off, in coast at 120 GeV/*c* (2009) [21]. The normalized transverse emittance had intentionally been blown up to 7.5 μm (twice the nominal).

While the compensation measurements reported so far were performed on the cycling machine at fairly low energy, in 2009 machine time was assigned for studies in coast at 120 GeV/*c*. Results are shown in Figs. 37 and 38.

Another measurement was performed one year later, in 2010, at 55 GeV/*c*. The results in Fig. 39 reveal that in this year and at this beam energy, the compensation was not as good as in the year before at 120 GeV/*c*. This could be due to some real energy dependence in the SPS (e.g. changes in field errors, or power-converter stability, etc.) or due to some other change in the machine between 2009 and 2010. Indeed, this and several other SPS beam studies in 2010 and 2011 noticed significant emittance growth (Fig. 40) and low lifetime in coast (Figs. 41 and 42), without wire excitation, at beam energies above the injection plateau (while in earlier years a poor beam lifetime had been noticed only at injection) [29].

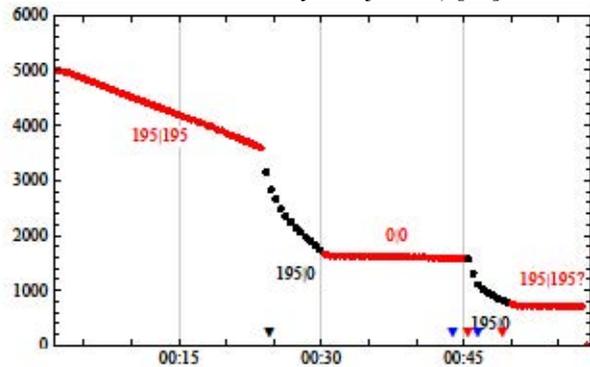

Figure 39: Beam intensity in units of $10^8$ protons as a function of time in hours while one or two wire compensators at 9.5σ distance were turned on (at 195 A, which for the given emittance value corresponded to the strength of long-range encounters for two LHC IPs with full beam intensity) and off, in coast at 55 GeV/*c* (2010) for the nominal LHC collision tune, $Q_x$=0.31, $Q_y$=0.32 [R. Calaga].

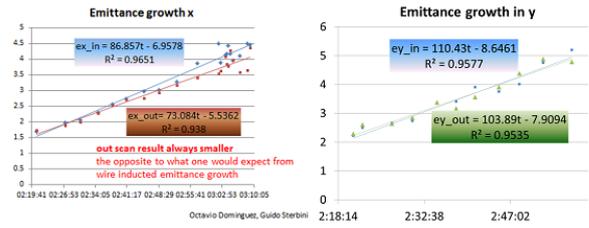

Figure 40: Emittance growth measured by (IN and OUT) wire scans during an SPS coast in 2010 without wire excitation [O. Dominguez, G. Sterbini].

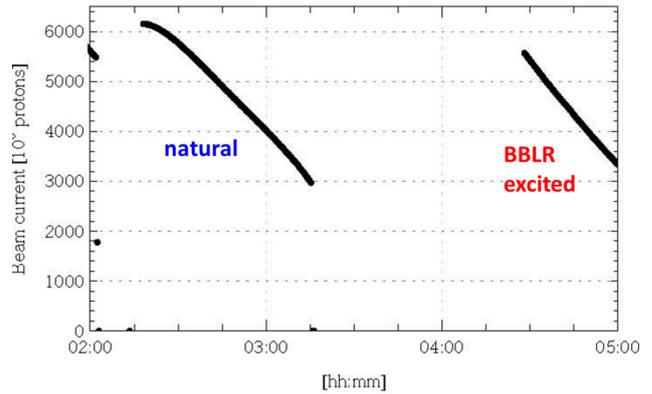

Figure 41: Beam intensity without and with single-wire excitation during an SPS coast in 2010 [O. Dominguez, G. Sterbini].

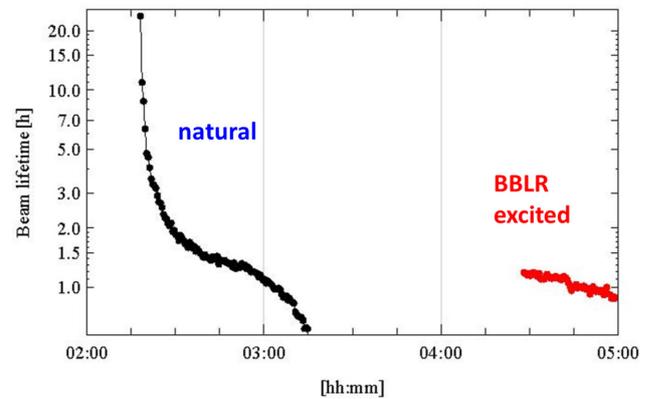

Figure 42: Instantaneous beam lifetime without and with single-wire excitation during an SPS coast in 2010, computed from the intensity data of Fig. 41 [O. Dominguez, G. Sterbini].

## ADVANCED BBLR STUDIES

Since different bunches along a train suffer a different number of long-range encounters (the so-called 'PACMAN effect' [1,2]) a dc wire compensator can never offer a perfect compensation. The left picture of Fig. 43 shows the ideal current pattern for PACMAN compensation and the right picture a schematic of an 'RF BBLR' based on a quarter-wave resonator [17, 18]. An experimental test set up (Figs. 44 and 45) has demonstrated the principle, with results as presented in Fig. 46.

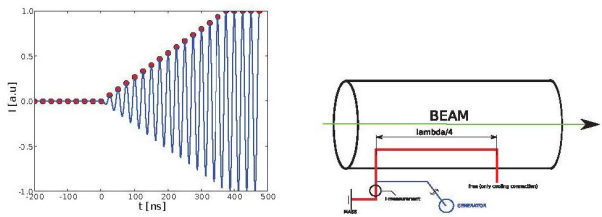

Figure 43: Ideal current pattern for compensation of individual long-range encounters with an amplitude-modulated 40-MHz signal (left) and schematic of an 'RF BBLR' built as a λ/4 resonator [19, 20, 30].

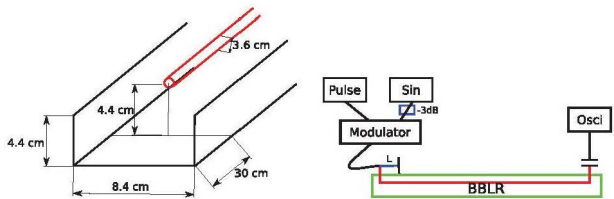

Figure 44: Drawings of 'RF BBLR' test set-up including some dimensions (left) and the cable length $L$ which can be changed for varying the coupling strength (right). Port 2 is connected capacitively in order not to modify the resonator properties [20, 30] [U. Dorda, F. Caspers, T. Kroyer].

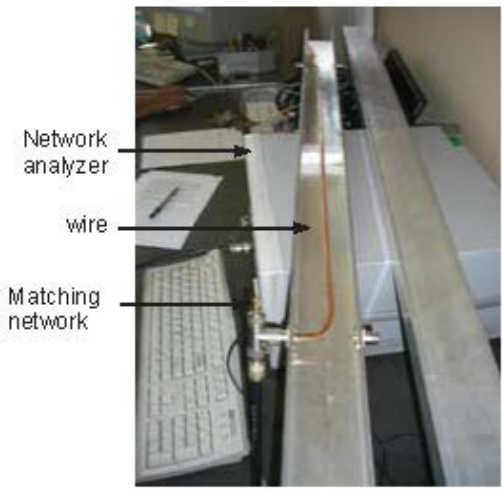

Figure 45: Photograph of 'RF BBLR' test set-up [20] [U. Dorda, F. Caspers, T. Kroyer].

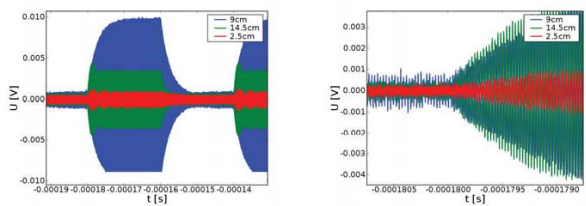

Figure 46: Test measurements showing the effect of varying coupling strength, i.e. the trade-off between rise time and gain: overview illustrating achievable resonator gains (left) and zoomed view showing the rise times for different couplings (right) [20] [U. Dorda, F. Caspers, T. Kroyer].

## CONCLUSIONS AND OUTLOOK

Ten years of pioneering wire-compensation studies at the CERN SPS taught many important lessons and gave rise to two PhDs (by G. Sterbini and U. Dorda).

Though the experimental conditions in the SPS were not always ideal (e.g. poor natural lifetime, short cycle times), the compensation of the first wire by a second wire always improved the beam lifetime significantly over a large range of parameters (current, distance, and tune). The results obtained confirm the simulations and strongly suggest that wire compensators will increase the operational flexibility and performance in the LHC.

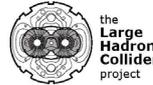

Figure 47: Space reservation for future LHC wire compensators made in 2004 [J.-P. Koutchouk].

For future wire BBLRs in the LHC, 3 m long sections were reserved in the LHC at 103.431 m to 106.431 m from the IP on either side of IP1 and IP5 (Fig. 47). This is close to the place where concrete shielding blocks have been installed (Fig. 48), the latter occupying the distance from 97.075 m to 100.225 m from the IP [31].

Recently, the two BNL wires stored at CERN were declared not useful for future SPS wire tests, as their beam pipe differs from the standard diameter [32].

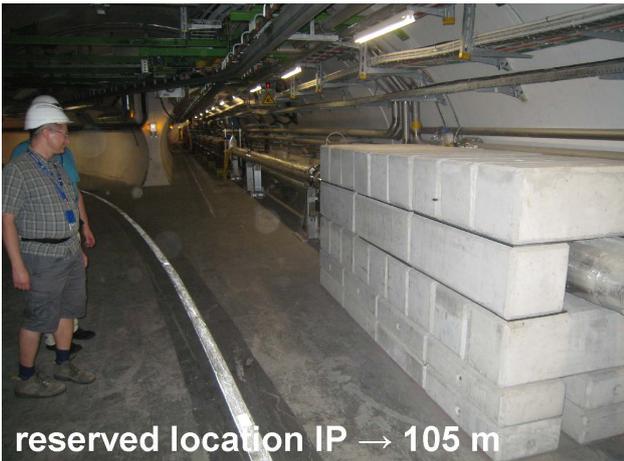

Figure 48: Concrete shielding block installed in the LHC tunnel close to the reserved wire-compensator location [photo R. Steinhagen].

## ACKNOWLEDGEMENTS

The reported SPS wire experiments would not have been possible without the ideas, help and important contributions of Gerard Burtin, Rama Calaga, Jackie Camas, Gijs de Rijk, Octavio Dominguez, Ulrich Dorda, Jean-Pierre Koutchouk, Elias Métral, Yannis Papaphilippou, Federico Roncarolo, Tanaji Sen, Vladimir Shiltsev, Guido Sterbini, Rogelio Tomas, and Jörg Wenninger.

## APPENDIX: CROSSING SCHEME STUDIES

The crossing scheme affects the extent of the tune footprint (Fig. 49) as well as the resonance excitation.

On 26 August 2004, an experiment on the crossing planes was conducted for the nominal emittance. Three configurations were implemented (see Fig. 50). Due to constraints imposed by the physical aperture and the different distances of the horizontal and vertical wire from the centre of the chamber (about 55 mm and 20 mm, respectively), a pure alternating crossing could not be realized. Instead a mixed scheme was chosen, modelling horizontal crossing at one wire and $45^\circ$ crossing at the other, by exciting both wires at the same current. Equal-plane crossings were modelled by exciting only one of the two wires at twice its original strength. For completeness, and to observe a larger effect on the beam lifetime, the first configuration was also tested at twice the strength, which simulates a two times higher beam intensity. The three configurations are shown in Fig. 50.

For all wire configurations, the beam lifetime was measured as a function of the vertical tune, which was varied between 0.26 and 0.33. Figure 52 shows the measurement results, which can be compared with the simulations presented in Fig. 51. Over most of the scanned tune range, the horizontal-horizontal crossing (BBLR2 excited at −240 A) exhibited the best beam lifetime, the pure $45^\circ$ crossing (BBLR1 at 240 A) the second best, and the mixed crossing (BBLR1 at +120 A, BBLR2 at -120 A) the lowest. At the two ends of the scan range, near the 7th and 3rd integer resonance, respectively, the pure $45^\circ$ crossing scheme was most robust, while for all others the lifetime strongly decreased here, possibly due to lattice nonlinearities. The lifetime without any wire excitation was comparable to that of the mixed-crossing case. Simulations and experiments are in reasonable agreement. In particular, the simulated diffusive aperture for pure *x-x* crossing is 10% larger than for *x*-$45^\circ$ or $45^\circ$-$45^\circ$ crossing, which appears consistent with the higher lifetime seen for this case. The larger variation with tune for the experimental data could be attributed to additional machine nonlinearities and/or perturbations, not included in the simulations.

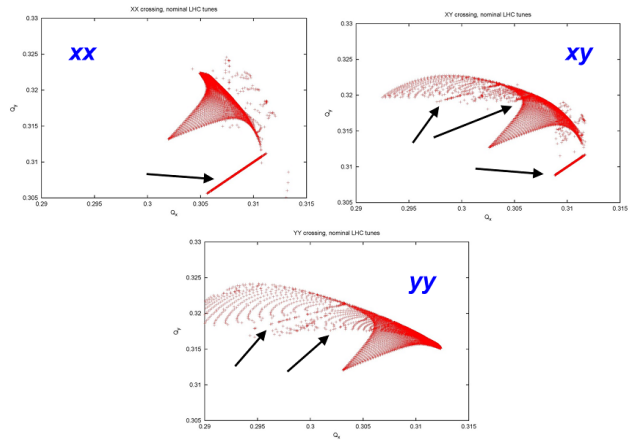

Figure 49: Simulated tune footprints for purely horizontal crossing, (top left), the nominal alternating horizontal-vertical crossing (top right) and purely vertical crossing (bottom) for regular bunches at the LHC collision tunes with collisions in and around the two main IPs. A few important resonances are highlighted by arrows.

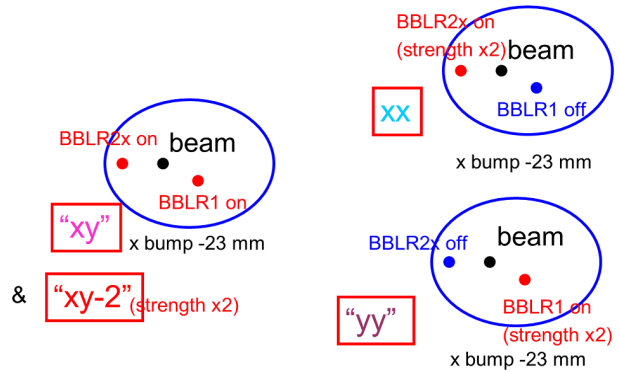

Figure 50: Approximations of different crossing schemes on 09/11/2004. The first configuration (left) models a mixed scheme with horizontal crossing at one wire and $45^\circ$ crossing at the other, the second (right top) a pure horizontal-horizontal crossing, and the third (right bottom) a pure $45^\circ$-$45^\circ$ crossing.

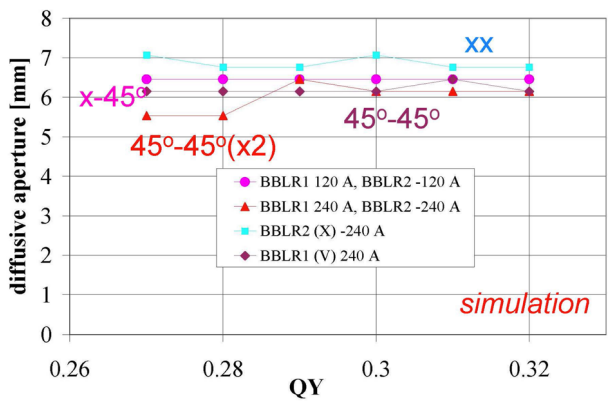

Figure 51: Dynamic aperture simulated by the WSDIFF code [34] (at $\beta \approx 50$ m) as a function of vertical tune keeping $Q_x = 0.31$, for the SPS wire configurations of Fig. 50.

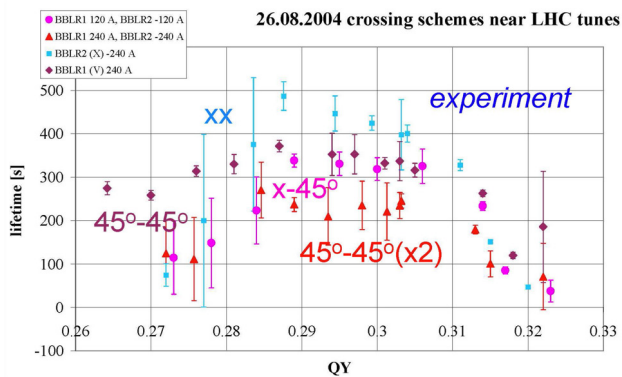

Figure 52: Beam lifetime measured as a function of the vertical tune for the three SPS wire configurations of Fig. 50. The horizontal tune was held constant at $Q_x \approx 0.31$.

On 9 November 2004, a second experiment was performed with reduced beam-wire distance and smaller emittance. One of the wires (BBLR2) had been rotated prior to this experiment, in order to allow for shorter transverse distances. The three configurations of Fig. 53 could then be realized. Again, it was not possible to implement a pure horizontal-vertical crossing. Instead a 45°-135° 'inclined hybrid crossing' [33] was modelled and its performance could be compared with that of a vertical-vertical or 45°-45° crossing.

Figure 54 displays the simulated dynamic aperture for these three configurations. The pure 45°-45° crossing has the smallest dynamic aperture. At vertical tunes of 0.29 or lower the vertical-vertical crossing is best, while at higher tunes the inclined-hybrid scheme yields the largest dynamic aperture. For completeness, the simulation results for a pure horizontal-vertical crossing are also indicated.

The measured beam lifetimes as a function of vertical tune are presented in Fig. 55. The lifetime was lowest for the 45°-45° crossing, the inclined hybrid crossing was best for tunes above 0.3, and the pure vertical-vertical crossing for lower tunes. All these results are consistent with the simulations in Fig. 54.

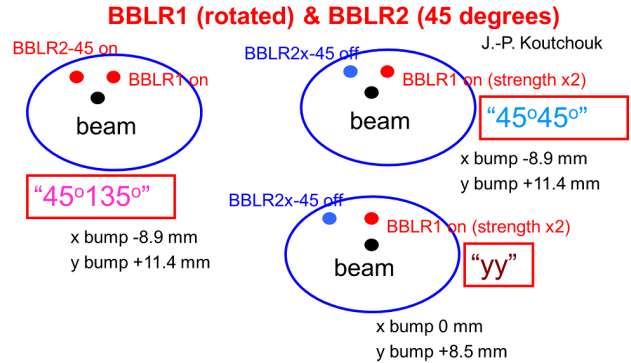

Figure 53: Approximations of different crossing schemes on 09/11/2004. The first configuration (left) models 45°-135° inclined hybrid collision [33], the second (right top) a double 45° hybrid crossing and the third (right bottom) a pure vertical-vertical crossing.

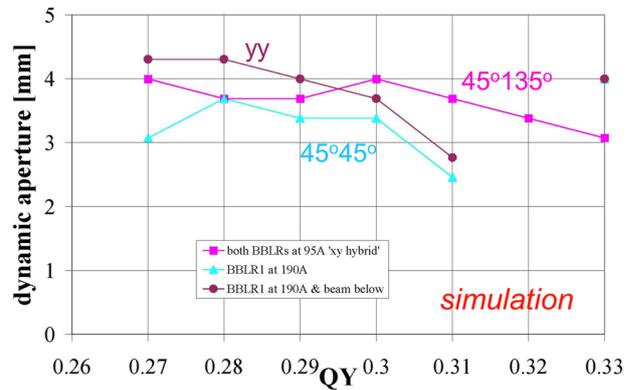

Figure 54: Dynamic aperture simulated by the WSDIFF code [34] (at $\beta \approx 50$ m) as a function of vertical tune keeping $Q_x = 0.31$, for the SPS wire configurations of Fig. 53.

Concluding this appendix, the beam lifetime was shown to vary with the crossing scheme. Experiments and simulations are mostly compatible.

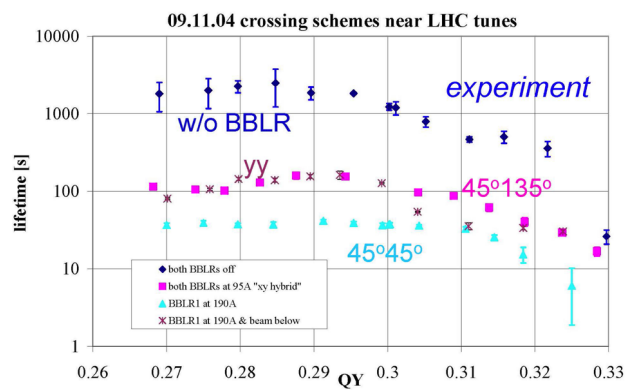

Figure 55: Beam lifetime measured as a function of the vertical tune for the three SPS wire configurations of Fig. 53. The horizontal tune was held constant at $Q_x \approx 0.31$.